\documentclass[a4paper,twoside,11pt,reqno]{article}
\usepackage[english]{babel}
\usepackage[dvips]{graphicx}
\usepackage{graphicx}
\usepackage[T1]{fontenc}
\usepackage[latin1]{inputenc}
\usepackage{amssymb,amsthm,dsfont,mathrsfs}
\usepackage{amsfonts,amsmath,euscript}
\usepackage{color}

\newtheorem{teo}{Theorem}[section]
\newtheorem{pro}[teo]{Proposition}

\newtheorem{rem}{Remark}[section]

\newcounter{example}[section]

           \addto{\captionsenglish}{}

\newcommand{\hs}{\hspace{3pt}}
\newcommand{\dst}{\displaystyle}
\newcommand{\dem}{{\bf Dem. }}
\newcommand{\fdem}{$\square$}

\newcommand{\sscr}{\scriptscriptstyle}

\newcommand{\nn}{\nonumber}

\newcommand{\titulo}[1]{\mbox{} \\ \noindent \textit{\textbf{\Large #1}}\\}

\newcommand{\autor}[1]{{ \textit{#1}  }}

\newcommand{\afil}[1]{{\small \noindent \textit{#1}}}

\renewcommand{\abstract}[1]{{\small \noindent \textbf{Abstract:} #1\\}}

\newcommand{\keywords}[1]{{\small \noindent \textbf{Keywords:} #1\\}}

\begin{document}

\begin{center}
\titulo{A new estimator for the tail-dependence coefficient}
\end{center}

\begin{center}
\autor{Marta Ferreira}

\afil{Department of Mathematics, University of Minho, Portugal\\
msferreira@math.uminho.pt\\}
\end{center}

\abstract{ Recently, the concept of tail dependence has been
discussed in financial applications related to market or credit
risk. The multivariate extreme value theory is a proper tool to
measure and model dependence, for example, of large loss events. A
common measure of tail dependence is given by the so-called
tail-dependence coefficient. We present a simple estimator of this
latter that avoids the drawbacks of the estimation procedure that
has been
used so far.
We prove strong consistency and asymptotic normality and analyze the
finite sample behavior through simulation. We illustrate with an
application to financial data.}


\section{Introduction}\label{sint}
Modern risk management is highly interested in assessing the amount
of extremal dependence, a growing phenomena in recent time periods
of volatile and bear markets. Correlation itself is not enough to
describe a tail dependence structure and often results in misleading
interpretations (see e.g. Embrechts \emph{et al.} \cite{emb+02},
2002 for examples).\\

 Multivariate  Extreme
value theory (EVT) is the natural tool to deal with tail dependence.
The so-called \emph{tail-dependence coefficient} (TDC)  has become a
popular measure in risk management. It is usually denoted $\lambda$
and measures the probability of occurring extreme values
 for one random variable (r.v.) given that
another assumes an extreme value too. More precisely,
\begin{eqnarray}\label{tdc}
\dst\lambda=\lim_{t\downarrow 0}P(F_X(X)>1-t|F_Y(Y)>1-t),
\end{eqnarray}
where $F_X$ and $F_Y$ are the distribution functions (d.f.'s) of
r.v.'s $X$ and $Y$, respectively. The TDC characterizes the
dependence in the tail of a random pair $(X,Y)$, in the sense that,
$\lambda>0$ corresponds to tail dependence whose degree is measured
by the value of $\lambda$, and $\lambda=0$ means tail independence.
Due to the emergent importance of this issue, it is not surprising
that the implementation of tail dependence measures and respective
estimation have attracted the attention of investigators. Sibuya
(\cite{sib}, 1960), Tiago de Oliveira (\cite{tiago}, 1962-63),
Ledford and Tawn (\cite{led+tawn1,led+tawn2}, 1996, 1997), Joe
(\cite{joe}, 1997), Coles \emph{et al.} (\cite{coles+}, 1999),
Embrechts \emph{et al.} (\cite{emb+03}, 2003), Frahm \emph{et al.}
(\cite{frahm+}, 2005), Schmidt and Stadtm\"{u}ller
(\cite{schm+stadt}, 2006), Ferreira and Ferreira (\cite{hf+mf1},
2011), are some references on this topic. Curiously, the first tail
dependence concept appearing in literature concerns the TDC, as far
back in the sixties with Sibuya (\cite{sib}, 1960). More precisely,
for Normal distributed random pairs, Sibuya (\cite{sib}, 1960) shows
that no matter how high we choose the correlation, if we go far
enough into the tail, extreme
events appear to occur independently in each margin. \\

Here we shall present an estimator for the TDC based on a new
procedure that avoids the main drawback of existing ones. Strong
consistency and asymptotic normality are stated and the finite
sample behavior is illustrated through simulation. An application to
financial data is presented at the end.

\section{EVT and tail dependence}\label{sevt+td}
The main objective of an extreme value analysis is to estimate the
probability of events that are more extreme than any that have
already been observed. The tools within EVT, in particular the use
of extremal models, enables extrapolations of this type. The central
result in univariate Extreme Value Theory (EVT) states that, for an
i.i.d.\hs sequence, $\{X_{n}\}_{\sscr n\geq 1}$, having common
distribution function (d.f.) $F$, if there are real constants
$a_{n}>0$ and $b_{n}$ such that,
\begin{eqnarray}\label{domatrac}
P(\max(X_{1},...,X_{n})\leq
a_{n}x+b_{n})=F^n(a_{n}x+b_{n})\longrightarrow_{n\rightarrow\infty} G(x)\, ,
\end{eqnarray}
for some non degenerate function $G_{\gamma}$, then it must be a
Generalized Extreme Value function (\textit{GEV}),
\begin{eqnarray}\label{gev}\nn
G(x)= \exp(-(1+\gamma x)^{-1/\gamma})\textrm{, }1+\gamma
x>0\textrm{, }\gamma \in \mathbb{R},
\end{eqnarray}
(for $\gamma=0$, $G(x)=\exp(-e^{-x})$) and we say that $F$ belongs
to the max-domain of attraction of $G$, in short, $F\in
\mathcal{D}(G)$. The parameter $\gamma$, known as the tail index, is
a shape parameter as it determines the tail behavior of $F$, being
so a crucial issue in EVT. More precisely, if $\gamma>0$ we are in
the domain of attraction Fréchet corresponding to a heavy tail,
$\gamma<0$ indicates the Weibull domain of attraction of light tails
and $\gamma=0$ means a Gumbel domain of attraction and an
exponential tail.

Models for dependent or non-identical distributed r.v.'s have also
been developed (see, for instance, Leadbetter \cite{lead+} 1983 and
Mejzler \cite{mejz} 1956, respectively).\\


In a multivariate framework, an extension of the univariate limiting
result in (\ref{domatrac}) is considered and the multivariate
maximum corresponds to the vector of component-wise maxima. Let
$\{\mathbf{X_n}=(X_{n,1},...,X_{n,d})\}_{n\geq 1}$ be an i.i.d.\hs
sequence of $d$-dimensional random vectors with common d.f.\hs $F$.
If there are real vectors of constants $\mathbf{b_{n}}$
 and positive $\mathbf{a_{n}}$ such that,
\begin{eqnarray}\label{mdomatrac}
P(\max(\mathbf{X_{1}},...,\mathbf{X_{n})}\leq
\mathbf{a_{n}x}+\mathbf{b_{n}})=F^n(\mathbf{a_{n}x}+\mathbf{b_{n}})
\longrightarrow_{n\rightarrow\infty} G(\mathbf{x})\, ,
\end{eqnarray}
for some non degenerate function $G$, then it must be
 a multivariate
extreme value distribution (MEV), given by
\begin{eqnarray}\label{mev}\nn
G(\mathbf{x})= \exp(-l(-\log G_1(x_1),...,-\log G_d(x_d)) ),
\end{eqnarray}
for some $d$-variate function $l$, where $G_j$, $j=1,...,d$, is the
marginal d.f. of $G$. We also say that $F$ belongs to the max-domain
of attraction of $G$, in short, $F\in \mathcal{D}(G)$. The function
$l$ in (\ref{mev}) is called \emph{stable tail dependence function}.

Let $F_j$ be the marginal d.f. of $F$. Since a sequence of random
vectors can only converge in distribution if the corresponding
marginal sequences do, we have, for $j=1,...,d$,
$$
F_j^n(a_{n,j}x_j+b_{n,j})\longrightarrow_{n\rightarrow\infty} G_{j}(x_j)
$$
Hence $G_{j}$ is a GEV and $F_j$ is in its domain of attraction.

In order to study the dependence structure of a MEV, it is
convenient to standardize the margins so that they are all the same.
A particular useful choice is the unit Fréchet, $\exp(-1/x)$
(observe that unit Fr\'{e}chet marginals can be obtained by
transformation $-1/\log F_j(X_j)$ for $j\in I\subset\{1,...,d\}$),
and
 the stable tail dependence function $l$ in (\ref{mev}) becomes
\begin{eqnarray}\label{l}
l(\mathbf{v})=-\log G(v_1^{-1},...,v_d^{-1})\textrm{, $\mathbf{v}\in\mathbf{[0,\infty)}$.}
\end{eqnarray}
Thus $l$ satisfies an important homogeneity property (of order $1$)
and hence, it is easy to establish that, except for the special case
of independence, all bivariate extreme value distributions(BEV) are
tail dependent ($\lambda>0$). Furthermore, we have
\begin{eqnarray}\label{lambda}
\lambda=2-l(1,1).
\end{eqnarray}

\textbf{Examples of parametric BEV models}

\begin{itemize}

\item \emph{Logistic:} $l(v_1,v_2)=(v_1^1/r+v_2^1/r)^r$, with
$v_j\geq 0$ and parameter $0<r\leq 1$; complete dependence is
obtained in the limit as $r\to 0$ and independence when $r = 1$.
\item \emph{Asymmetric Logistic:} $l(v_1,v_2)=(1-t_1)v_1+(1-t_2)v_2+((t_1v_1)^1/r+(t_2v_2)^1/r)^r$, with
$v_j\geq 0$ and parameters $0<r\leq 1$ and $0\leq t_j\leq 1$,
j=1,2; when $t_1 = t_2 = 1$ the asymmetric logistic model is
equivalent to the logistic model; independence is obtained when
either $r = 1$, $t_1 = 0$ or $t_2 = 0$. Complete dependence is
obtained in the limit when $t_1 = t_2 = 1$ and $r$ approaches
zero.
\item \emph{H\"{u}sler-Reiss:} $l(v_1,v_2)=v_1\Phi(r^{-1}+\frac{1}{2}r\log(v_1/v_2))+
v_2\Phi(r^{-1}+\frac{1}{2}r\log(v_2/v_1))$, with parameter $r>0$
and where $\Phi$ is the standard normal d.f.; complete
dependence is obtained as $r\to \infty $ and independence as
$r\to 0 $.

\end{itemize}

\section{Estimation}\label{sestim}
The $d$-variate stable tail dependence function in (\ref{l}) can
also be formulated as
\begin{eqnarray}\label{stfmulti}
\lim_{t\to\infty}tP\Big(F_1(X_1)>1-\frac{x_1}{t} \vee ... \vee F_d(X_d)>1-\frac{x_d}{t}\Big)
\end{eqnarray}
since, by applying the unit Fréchet marginals, we have successively,
\begin{eqnarray}\label{stfmulti}
\begin{array}{rl}
&\lim_{t\to\infty}tP\Big(F_1(X_1)>1-\frac{x_1}{t} \vee ... \vee F_d(X_d)>1-\frac{x_d}{t}\Big)\\
=& \lim_{t\to\infty}tP\Big(X_1>\frac{t}{x_1} \vee  ... \vee X_d>\frac{t}{x_d}\Big)\\
=& \lim_{t\to\infty}tP\Big(X_1>\frac{t}{x_1} \vee ... \vee X_d>\frac{t}{x_d}\Big)\\
=& \lim_{t\to\infty}t\Big(1-F\Big(\frac{t}{x_1} ,..., \frac{t}{x_d}\Big)\Big)\\
=& \lim_{t\to\infty}-\log F^t\Big(\frac{t}{x_1} ,..., \frac{t}{x_d}\Big) \\
=&-\log G\Big(\frac{1}{x_1} ,..., \frac{1}{x_d}\Big) .
\end{array}
\end{eqnarray}
Therefore, based on (\ref{lambda}) and (\ref{stfmulti}), the TDC in
(\ref{tdc}) can be stated like follows:
\begin{eqnarray}\label{lambdaH}
\lambda=2-\lim_{t\to\infty}tP\Big(F_1(X_1)>1-\frac{1}{t} \vee F_2(X_2)>1-\frac{1}{t}\Big).
\end{eqnarray}

Huang (1992 \cite{huang}), considered the estimator based on
(\ref{lambdaH}) by plugging-in the respective empirical
counterparts,
\begin{eqnarray}\label{estimH}
\widehat{\lambda}^{(H)}=2-\frac{1}{k_n}\sum_{i=1}^n\mathbf{1}_{\{\widehat{F}_1(X_1)>1-\frac{k_n}{n} \vee \widehat{F}_2(X_2)>1-\frac{k_n}{n}\}},
\end{eqnarray}
where  $\widehat{F}_j$ is the empirical d.f.\hs of $F_j$, $j=1,2$.
Concerning estimation accuracy, some  modifications of this latter
may be used, like replacing the denominator $n$ by $n+1$, i.e.,
considering
$$
\widehat{F}_j(u)=\frac{1}{n+1}\sum_{k=1}^n\mathbf{1}_{\{X_j^{(k)}\leq
u\}}
$$
(for a discussion on this topic see, for instance, Beirlant et al.
\cite{beirl+} 2004). The consistency and asymptotic normality of
estimator $\widehat{\lambda}^{(H)}$ is derived under the condition
that $\{k_n\}$ is an intermediate sequence, i.e., $k_n\to\infty$ and
$k_n/n\to 0$, as $n\to\infty$.
 The choose of the value $k$ in the sequence $\{k_n\}$ that allows
the better trade-off between bias and variance is of major
difficulty, since small values of $k$ come along with a large
variance whenever an increasing $k$ results in a strong bias.
Therefore, simulation studies have been carried out in order to find
the best value of $k$ that allows this compromise. 
The other estimators, either for the stable tail dependence function
$l$ or for the TDC,  that have been considered in literature (for a
survey, see for instance, respectively, Krajina  \cite{krajina} 2010
and Frahm \emph{et al.} \cite{frahm+} 2005) are also based on
asymptotic results with the same drawback of including an
intermediate sequence, already referred above.

The approach that is presented here avoid this problem. It is based
on an estimation procedure for the stable tail dependence function
only involving a sample mean. More precisely, by Proposition 3.1 in
Ferreira and Ferreira (\cite{hf+mf2}, 2011), we have that
\begin{eqnarray}\label{l2}
l(x_1,x_2)=\frac{E(F_1(X_1)^{1/x_1}\vee F_2(X_2))^{1/x_2}}{1-E(F_1(X_1)^{1/x_1}\vee F_2(X_2))^{1/x_2}}.
\end{eqnarray}
Therefore, based on (\ref{lambda}) and (\ref{l2}) we propose
estimator
\begin{eqnarray}\label{estimnew}
\widehat{\lambda}=3-(1-\overline{\widehat{F}_1(X_1)\vee \widehat{F}_2(X_2)})^{-1}
\end{eqnarray}
where $\overline{\widehat{F}_1(X_1)\vee \widehat{F}_2(X_2)}$ is the
sample mean of $\widehat{F}_1(X_1)\vee \widehat{F}_2(X_2)$, i.e.,
\begin{eqnarray}\nn
\begin{array}{c}
\dst\overline{\widehat{F}_1(X_1)\vee \widehat{F}_2(X_2)}=\frac{1}{n}\sum_{i=1}^n\big[\widehat{F}_1(X_1^{(i)})\vee \widehat{F}_2(X_2^{(i)})\big].
\end{array}
\end{eqnarray}
\begin{pro}
Estimator $\widehat{\lambda}$ in (\ref{estimnew}) is asymptotically
normal whenever $F$ has continuous marginals and continuous partial
derivatives. Moreover it is strong consistent.
\end{pro}
\dem By Fermanian \emph{et al.} (\cite{ferm}, 2002, Theorem 6), we
have that
$$
\frac{1}{\sqrt{n}}\sum_{i=1}^n \{J(\widehat{F}_1(X^{(i)}_1),\widehat{F}_2(X^{(i)}_2))
-E(J(\widehat{F}_1(X^{(i)}_1),\widehat{F}_2(X^{(i)}_2)))\}\to\int_{[0,1]^d}\mathbb{G}(u_1,u_2)dJ(u_1,u_2)
$$
in distribution in $\ell^\infty([0,1]^2)$, where the limiting
process and $\mathbb{G}$ are centered Gaussian, and
$J(u_1,u_2)=\max(u_1,u_2)$, $(u_1,u_2)\in[0,1]^2$. The asymptotic
normality is now derived from a general version of the Delta Method
as considered in Schmidt and Stadtm\"{u}ller \cite{schm+stadt}
(2006; Theorem 13).

The strong consistency is straightforward from  Proposition 3.7 in
Ferreira and Ferreira (\cite{hf+mf2}, 2011). \fdem\\

\begin{rem}
Observe that, if the marginals $F_j$, $j=1,2$, in (\ref{estimnew})
are known, the asymptotic normality of $\widehat{\lambda}$ is
straightforward by the Central Limit Theorem and the usual Delta
Method. More precisely,
\begin{eqnarray}\label{normxepsilon1}
\sqrt{n}(\widehat{\lambda}-\lambda)\to
N(0,\sigma^2),
\end{eqnarray}
where
\begin{eqnarray}\nn
\begin{array}{c}
\sigma^2
=\frac{l(1,1)\big(1+l(1,1)\big)^2}{\big(2+l(1,1)\big)}.
\end{array}
\end{eqnarray}
Details of the calculations can be seen in Proposition 3.3 of
Ferreira and Ferreira (\cite{hf+mf2}, 2011). The strong consistency
is also immediately derived from the sample mean.
\end{rem}

\subsection{Simulations}

We consider $1000$ independent copies of $n = 50,100,500,1000$
i.i.d.\hs pseudo-random vectors generated from three different BEV
models considered in Section \ref{sevt+td}: logistic, asymmetric
logistic and H\"{u}sler-Reiss. We estimate the TDC through our
estimator ($\widehat{\lambda}$). For comparison, we also compute
estimator $\widehat{\lambda}^{(H)}$ and, the required choice of $k$
to balance the variance-bias problem is based on the procedure in
Schmidt and Stadtm\"{u}ller (\cite{schm+stadt}, 2006). The empirical
bias and the root mean-squared error (rmse) for all implemented TDC
estimations are derived and presented in Table \ref{tab1}. Our
estimator $\widehat{\lambda}$ clearly outperforms estimator
$\widehat{\lambda}^{(H)}$.

\begin{table}
\caption{Sample absolute bias and RMSE of the non-parametric TDC
estimators $\widehat{\lambda}$ and $\widehat{\lambda}^{(H)}$ for BEV
Logistic, Asymmetric Logistic and H\"{u}sler-Reiss
models.\label{tab1}}\vspace{0.5cm}
\begin{tabular}{ccc}
Logistic ($r=0.7$)&&    \\
 \hline \hline  \\
$\lambda=0.3755$& $\widehat{\lambda}$& $\widehat{\lambda}^{(H)}$\\
 \hline

  &bias (rmse)& bias (rmse)   \\
  (n=50)&0.0019 (0.0994)& 0.0395 (0.1962)\\
  (n=100)&0.0052 (0.0711)&0.0389 (0.1412)  \\
   (n=500)&0.0006 (0.0330)& 0.0216 (0.0883)  \\
  (n=1000)&0.0002 (0.0232)& 0.0099 (0.1379) \\

 \hline\\

 \\

Asym. Logistic ($r=0.7,t_1=t_2=0.5$)&&\\
 \hline\hline\\

  $\lambda=0.1877$& $\widehat{\lambda}$& $\widehat{\lambda}^{(H)}$\\
  \hline
    &bias (rmse)& bias (rmse)   \\
  (n=50)&0.0085 (0.1147)& 0.0527 (0.1836)\\
  (n=100)&0.0053 (0.0824)&0.0635 (0.1363)  \\
   (n=500)&0.0020 (0.0389)& 0.0335 (0.0847)  \\
  (n=1000)&0.0014 (0.0287)& 0.0038 (0.1193) \\
\hline\\

\\
H\"{u}sler-Reiss($r=0.7$)&&\\
 \hline\hline\\
   $\lambda=0.1531$& $\widehat{\lambda}$& $\widehat{\lambda}^{(H)}$\\
 \hline

  &bias (rmse)& bias (rmse)   \\
  (n=50)&0.0119 (0.1293)& 0.0729 (0.1893)\\
  (n=100)&0.0077 (0.0838)&0.0706 (0.1387)  \\
   (n=500)&0.0020 (0.0383)& 0.0378 (0.0851)  \\
  (n=1000)&0.0020 (0.0293)& 0.0060 (0.1084) \\
\hline
\end{tabular}
\end{table}

\subsection{Application to financial data}
\begin{figure}[!htb]
\begin{center}
\begin{tabular}{c}
\includegraphics[width=5cm,height=4.55cm]{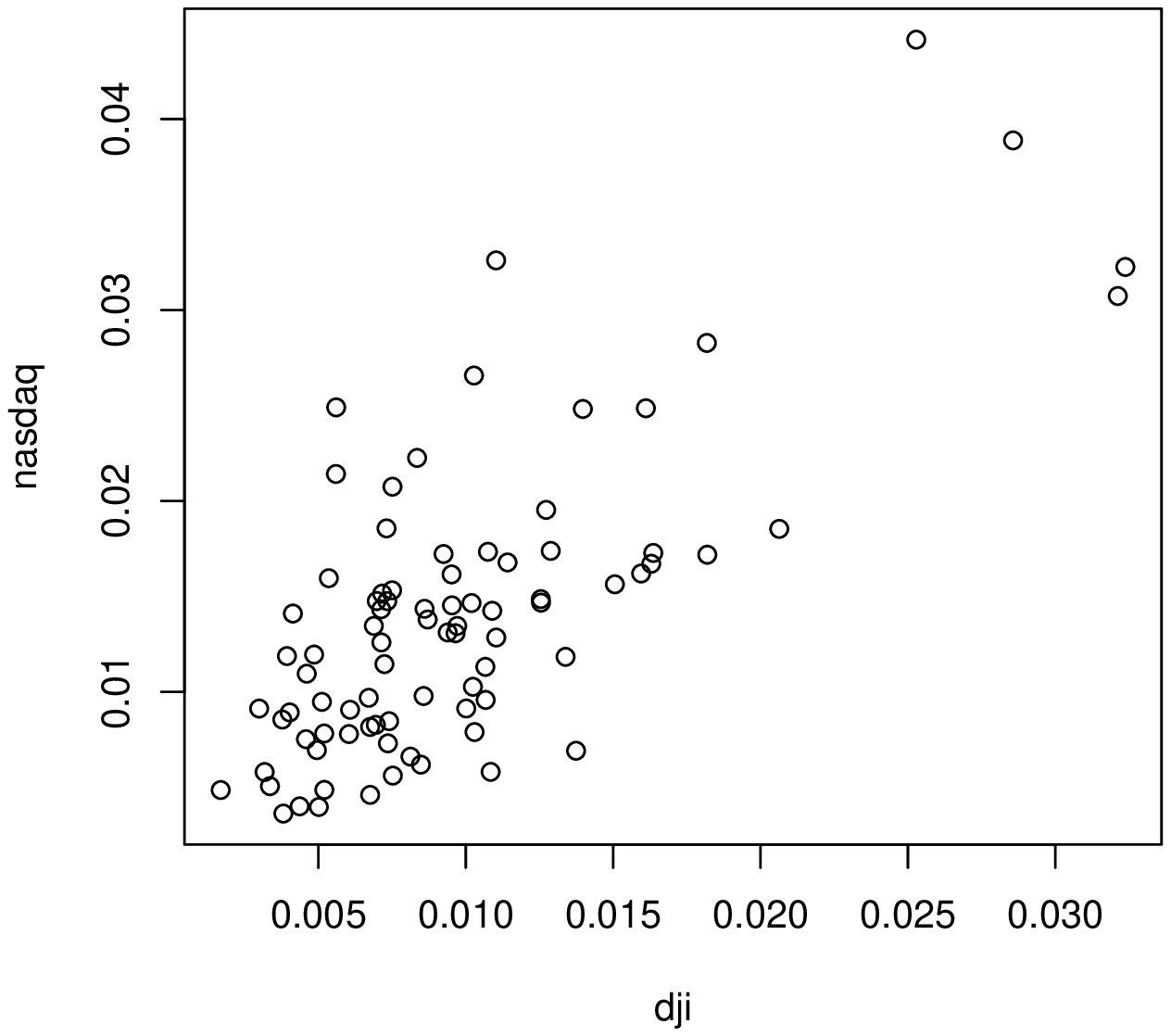}
\includegraphics[width=5cm,height=4.55cm]{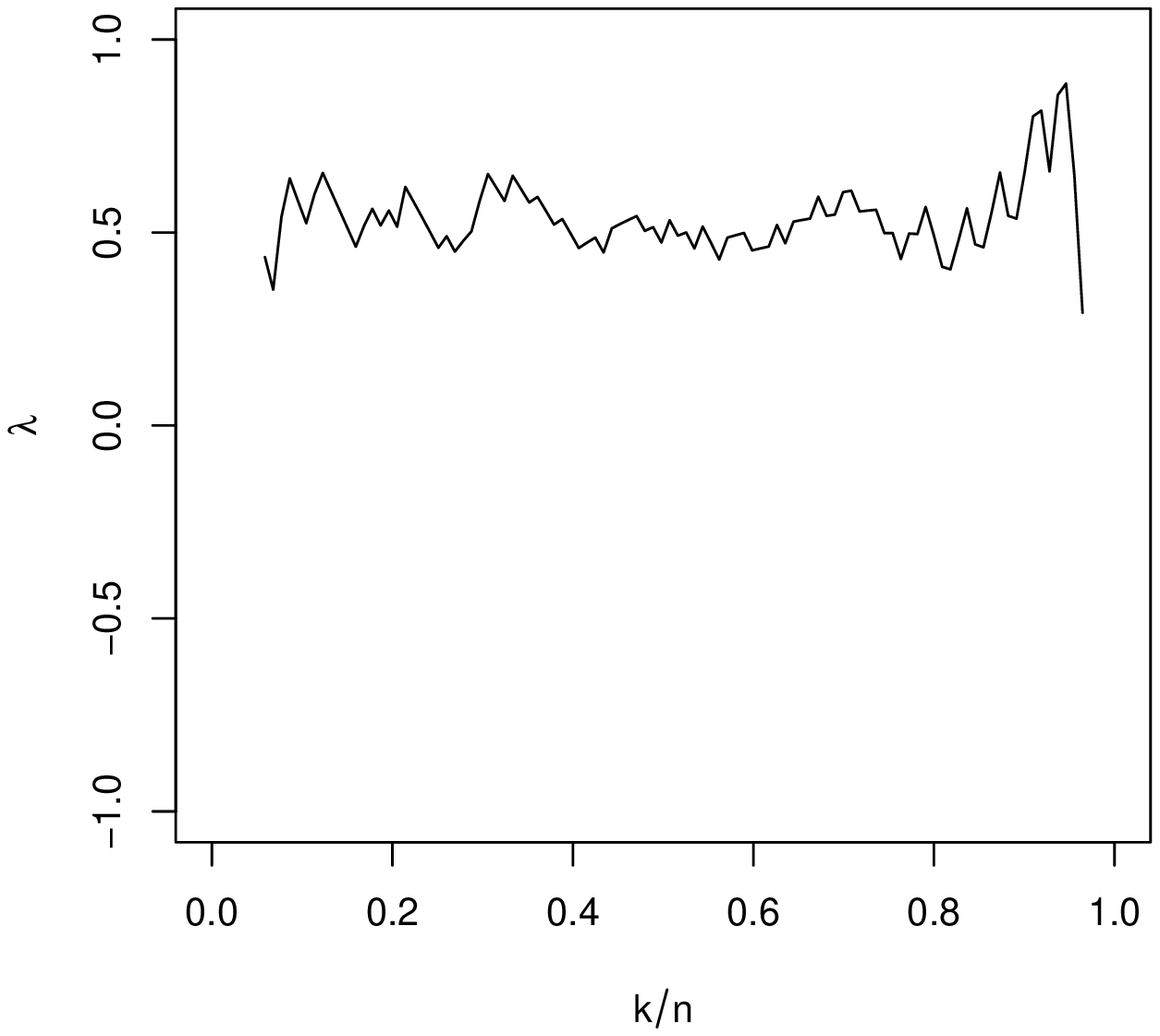}
\end{tabular}
\caption{Scatter plot of Dow
Jones versus NASDAQ monthly maximum negative log-returns ($n=84$ data points)
and the corresponding TDC estimates $\widehat{\lambda}^{(H)}$ for various $k/n$. \label{fig1}}
\end{center}
\end{figure}

We shall see evidence of tail dependence in financial data. We
consider the monthly maximum of the negative log-returns of Dow
Jones and NASDAQ index for the time period 1994-2004. The
corresponding scatter plot and TDC estimate plot of
$\widehat{\lambda}^{(H)}$ for various $k$ (Figure \ref{fig1}) show
the presence of tail dependence and the order of magnitude of the
tail-dependence coefficient. Moreover, the typical variance-bias
problem for various threshold values $k$ can be observed, too. In
particular, a small k induces a large variance, whereas an
increasing k generates a strong bias of the TDC estimate. The
threshold choosing procedure of $k$ used in Section \ref{sestim}
leads to a TDC estimate of $\widehat{\lambda}^{(H)}=0.5556$ and from
our estimator we derive $\widehat{\lambda}=0.5268$.


\begin{thebibliography}{99}

\bibitem{beirl+} Beirlant, J.,  Goegebeur, Y., Segers, J. e Teugels, J. (2004). \emph{Statistics of Extremes: Theory and Application.} John Wiley.

\bibitem{coles+} Coles, S., Heffernan, J., Tawn, J. (1999). Dependence measures
for extreme value analysis, Extremes 2 339-366.

\bibitem{emb+02} Embrechts, P., McNeil, A., Straumann, D. (2002).  Correlation and
dependence in risk management: properties and pitfalls In: Risk
Management: Value at Risk and Beyond, ed. M.A.H. Dempster, Cambridge
University Press, Cambridge, pp. 176-223.

\bibitem{emb+03} Embrechts, P., Lindskog, F., McNeil, A. (2003). Modelling
Dependence with Copulas and Applications to Risk Management, In:
Handbook of Heavy Tailed Distibutions in Finance, ed. S. Rachev,
Elsevier, Chapter 8: 329-384.

\bibitem{ferm}Fermanian, J.-D., Radulovi\'{c}, D. (2004).  Wegkamp, M., Weak
convergence of empirical copula processes. Bernoulli 10(5) 847-860.


\bibitem{hf+mf1}Ferreira, H. , Ferreira M.. Tail dependence between order statistics, Journal of Multivariate Analysis (2011),
doi:10.1016/j.jmva.2011.09.001.

\bibitem{hf+mf2}Ferreira, H., Ferreira  M., Extremal dependence: some
contributions (Submitted) (arXiv:1108.1972v1).

\bibitem{frahm+} Frahm, G., Junker,  M., Schmidt R. (2005).
Estimating the tail-dependence coefficient: properties and pitfalls.
Insurance Math. Econom.  37 (1) 80-100.


\bibitem{huang} Huang, X. (1992). Statistics of Bivariate Extreme Values. Ph. D.
thesis, Tinbergen Institute Research Series 22, Erasmus University
Rotterdam.

\bibitem{joe} Joe, H. (1997). Multivariate Models and Dependence Concepts, Chapman \& Hall,
London.

\bibitem{krajina} Krajina, A. (2010). \emph{An M-Estimator of Multivariate Tail Dependence}.
Tilburg: Tilburg University Press.

\bibitem{lead+} {Leadbetter, M.\hs R., Lindgren, G.\hs e Rootz\'{e}n, H.} (1983). {\em Extremes and Related Properties of Random Sequences and
Processes}. Springer-Verlag, New-York, Heidelberg-Berlin.

\bibitem{led+tawn1} Ledford, A., Tawn, J.A. (1996). Statistics for near independence in
multivariate extreme values, Biometrika 83 169-187.

\bibitem{led+tawn2}Ledford, A. Tawn, J.A. (1997). Modelling dependence within joint tail
regions, J. R. Stat. Soc. Ser. B Stat. Methodol. 59 475-499.


\bibitem{mejz} Mejzler, D. (1956).  On the problem of the limit distribution for the maximal
term of a variacional series, L'vov Politechn. Inst. Naucn. ZP., 38
90-109.



\bibitem{schm+stadt}Schmidt, R., Stadtmüller, U.  (2006). Nonparametric estimation of tail
dependence, Scandinavian J. Statist. 33 307-335.

\bibitem{sib} Sibuya, M. (1960). Bivariate extreme statistics, Ann. Inst. Statist. Math.
11 195-210.


\bibitem{tiago} Tiago de Oliveira, J. (1962-1963). Structure theory of bivariate extremes,
extensions, Est. Mat. Estat. Econ. 7  165-195.



\end{thebibliography}
\end{document}